\documentclass[11pt,a4paper]{article}

\usepackage{graphicx}

\begin{document}
\pagestyle{plain}
\begin{center}
{\large \bf Supplementary material for:} \\
\vspace{1cm}
{\Large \bf Multiscale non-adiabatic dynamics with radiative decay, case study on the post-ionization fragmentation of
rare-gas tetramers.} 
\vspace{0.5cm}

Ivan Jane\v{c}ek \\
{\small \it Institute of Geonics of the AS CR, v.v.i.,  \&
Institute of Clean Technologies for Mining and Utilization of Raw Materials for Energy Use, Studentsk\'{a} 1768, 708~00 Ostrava, Czech Republic}\\
\vspace{0.5cm}
Tom\'{a}\v{s} Jan\v{c}a, Pavel Naar \\
{\it \small Department of Physics, Faculty of Sciences, University of Ostrava, 30. dubna 22, 701~03 Ostrava, Czech Republic}\\
\vspace{0.5cm}
Frederic Renard \\
{\it \small Facult\'{e} des Sciences, Universit\'{e} du Maine, 72085 Le Mans Cedex 9, France}\\
\vspace{0.5cm}
Ren\'{e} Kalus \\
{\it \small Centre of Excellence IT4Innovations \& Department of Applied Mathematics,
V\v{S}B - Technical University of Ostrava, 17. listopadu 15, 708 33 Ostrava, Czech Republic}\\
\vspace{0.5cm}
Florent X. Gad\'{e}a \\
{\it \small LCPQ and UMR5626 du CNRS, IRSAMC, Universit\'{e} de Toulouse, 118 route de Narbonne, 31062 Toulouse Cedex, France}
\end{center}
\vspace{2cm}
%
\begin{center}
{\small \bf Abstract}
\end{center}
{\small In this supplementary material, we recollect, for reader's convenience, the general scheme of suggested multiscale model (Sec. \ref{scheme}), and basic informations about approaches used for pilot study: a detailed description of the interaction model
(Sec. \ref{interact}) and dynamical methods used for the dark dynamics step (Sec. \ref{non_rad_dyn}) reported
previously in two preceding studies \cite{janecek06,janecek09}.
In addition, a detailed description of the treatment of radiative processes is also given (Sec. \ref{rad_dyn}).}
\vspace{1cm} \\
\noindent Last update: \today

\newpage
\setlength{\footskip}{15mm}
\setlength{\topmargin}{0mm}
\setlength{\headheight}{0mm}
\setlength{\headsep}{0mm}
%

%
\section{General model scheme}
\label{scheme}
%

\includegraphics [scale=0.58]{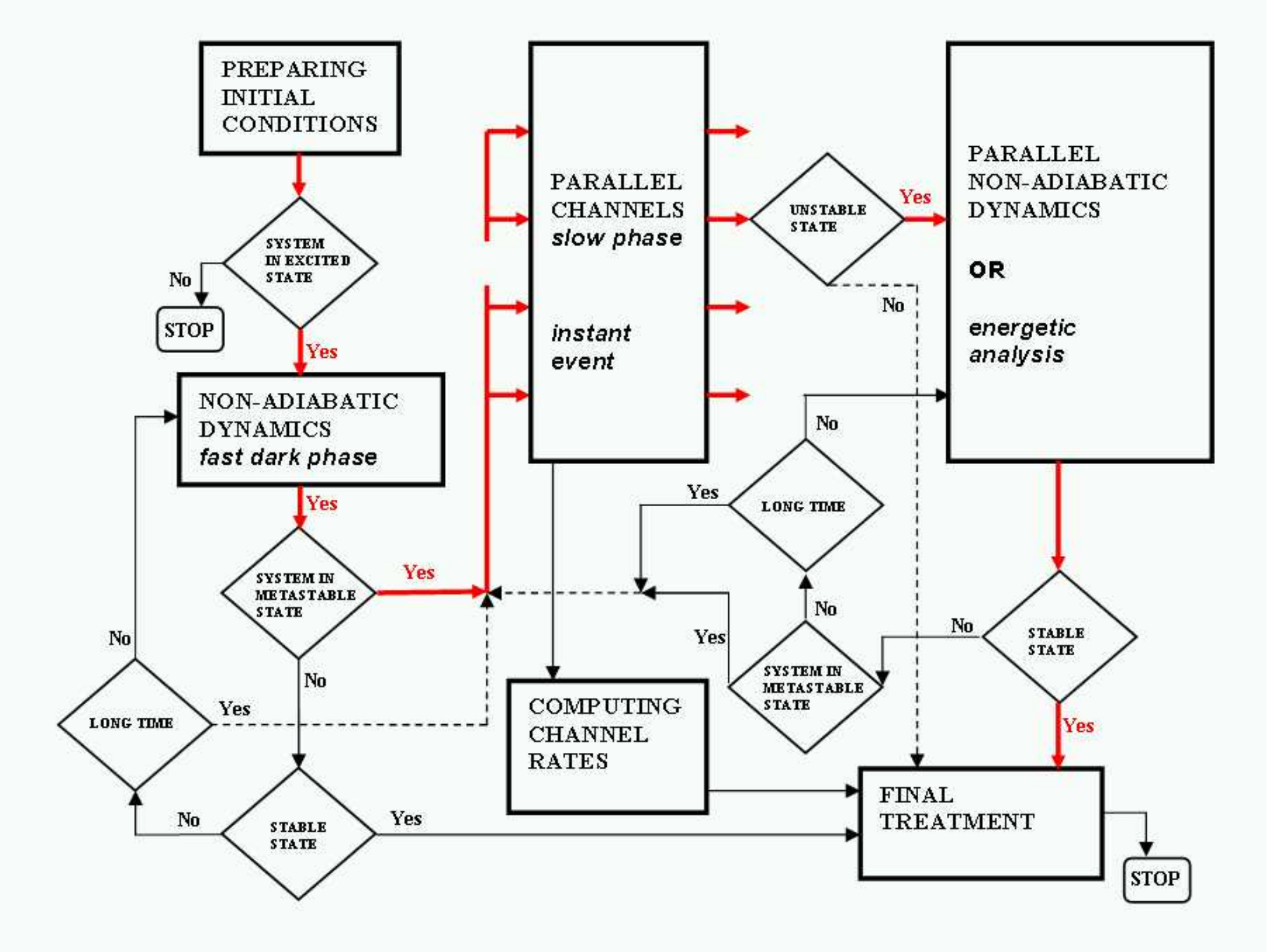}
\\ \textbf{Figure:}{ General scheme of suggested multiscale model.}

%
\section{Interaction model}
\label{interact}
The intra-cluster interactions are described within an extended {\it diatomics-in-molecules} (DIM) model
with the spin-orbit interaction included.
The DIM approach was developed by Ellison \cite{ellison63} and later on applied to singly ionized
rare-gas cluster cations by Kuntz and Valldorf \cite{kuntz88}. How the spin-orbit (SO) interaction can be included
in the DIM model was proposed by Amarouche et al. \cite{amarouche88}. Note that, in addition to the SO term, further extensions
to the original DIM approach can also be considered, e.g., the inclusion of the leading three-body polarization forces \cite{amarouche88}.
However, we do not use these extensions in the present study as they do not contribute much for small clusters \cite{janecek06},
and, consequently, we omit them from the following explanation.

The original DIM approach consists in a) re-writing the electronic hamiltonian
as a sum of diatomic and atomic contributions,
\begin{equation}\label{dim}
\mathrm{\hat{H}} = \sum_{j=1}^{n-1}\sum_{k=j+1}^n \mathrm{\hat{H}}_{jk}-(n-2)\sum_{k=1}^n \mathrm{\hat{H}}_k,
\end{equation}
where $n$ denotes the number of atoms, and b) designing an appropriate basis
set of wave functions for which the elements of the corresponding hamiltonian matrix
can be calculated by means of the electronic energies of atomic and diatomic fragments.
If the SO coupling is not considered, the atomic contributions
of Eq. \ref{dim} are constant and their sum can be identified with the zero energy level.\footnote{In our work, 
the zero of energy is identified with the energy of fully dissociate state,
$\mathrm{Rg}^+ + (n-1)\mathrm{Rg}$ , calculated with the SO interaction \emph{not} included and
the atomic contributions of Eq. \ref{dim} can be omitted.
If, on the other hand, the SO coupling is considered, the atomic contribution corresponds either to the energy of
$\mathrm{Rg}^+(^2\mathrm{P}_{3/2})$ or $\mathrm{Rg}^+(^2\mathrm{P}_{1/2})$ measured from the 
SO-free $\mathrm{Rg}^+$ level.
}
The diatomic energies are to be supplied from independent sources (usually from
{\it ab initio} calculations).

The basis set proposed for an $n$-atom rare-gas cluster cation, Rg$_n^+$, 
consists for the SO-free model of $3n$ valence-bond
Slater determinants, $|\Phi_{k,p_m} \rangle$ (where $k = 1,\cdots,n$ and $m = x,y,z$) and
represents states with the positive charge localized in a valence $p_m$-orbital of atom $k$. The corresponding
$3n \times 3n$ hamiltonian matrix,
\begin{equation}\label{dim_mtrx}
H_{k,p_m;k',p_m'} \equiv \langle \Phi_{k,p_m} |\mathrm{\hat{H}}| \Phi_{k',p'_m} \rangle,
\end{equation}
is constructed as described in Ref. \cite{kuntz88} from diatomic potential energy curves for the electronic ground
state of the neutral dimer, Rg$_2$, and the electronic ground and three lowest excited states of the ionic
dimer, Rg$_2^+$. In our calculations, we have used semiempirical curves for neutral dimers \cite{azizKr,azizXe} and accurate
{\it ab initio} curves for ionic diatoms \cite{krypton1,gapaxe}.

This simple model
is easily augmented \cite{amarouche88} with SO coupling terms via a semi-empirical {\it atoms-in-molecules} scheme
\cite{cohen}. If the SO coupling is taken into account, the number of
the basis set wave functions doubles,
as there are two possible orientations of the spin of the electron removed from the valence shell of a particular atom,
$s_z = \pm 1/2$, as well as the dimension of the electronic hamiltonian matrix. In addition,
the matrix of Eq. \ref{dim_mtrx} must be replaced by ($\delta$ denotes the Kronecker delta) \cite{amarouche88}
\begin{equation}\label{dimso_mtrx}
H_{k,p_m,s_z;k',p_m',s'_z}^\mathrm{(SO)} = H_{k,p_m;k',p_m'}\delta_{s_z;s_z'} + h_{p_m,s_z;p_m',s_z'}^\mathrm{(SO)}\delta_{k;k'},
\end{equation}
where
\begin{equation}\label{dimso_hmtrx}
h_{p_m\sigma,l\sigma}^\mathrm{(SO)}\delta_{k;k'} = \xi \langle \phi_{k,p_m,s_z}|\widehat{L}_k\widehat{s}_k|\phi_{k',p_m',s_z'} \rangle,
\end{equation}
$\xi$ is the~SO coupling constant,
and  $\widehat{L}_k$ and $\widehat{s}_k$
are angular and spin operators for~the~$k$-th atom, respectively. The SO constants are independent inputs
to the model and have been extracted here 
from experiments reporting on the SO splitting between the $^2P_{1/2}$ and  $^2P_{3/2}$ states
of atomic monomers \cite{nist}. Hereafter, we denote this extended DIM model by DIM+SO.

In the following text we use a simplified indices for the electronic wave function and hamiltonian matrix components, e.g.,
$\alpha = [k,p_m,s_z]$ etc.
%
%
\section{Non-radiative dynamics}
\label{non_rad_dyn}
The semi-classical dynamical method (classical nuclei and quantum electrons)
we use in our work for the non-radiative stage of our calculations,
the MFQ-AMP/S method of Ref. \cite{janecek09}, combines a) the well known Ehrenfest mean-field
approach \cite{ehrenfest27}, detailed for the rare-gas cluster cations in \cite{janecek06},
with b) the inclusion of quantum decoherence as introduced
in Ref. \cite{janecek09}.
\subsection{Mean-field method}
The equations of motion for a system of classical nuclei surrounded by a cloud of electrons
can be written within the mean-field approximation as coupled classical Hamilton equations for the nuclei
\begin{equation}\label{d_ham_q}
\dot{q}_i = \frac{p_i}{m_i},
\end{equation}
\begin{equation}\label{d_ham_p}
\dot{p_i} = \langle \psi |-\frac{\partial \mathrm{\hat{H}}}{\partial q_i}| \psi \rangle
\end{equation}
and time dependent Schr\"{o}dinger equation for the electrons
\begin{equation}\label{d_schr}
i \hbar \frac{\partial | \psi \rangle}{\partial t} =  \mathrm{\hat{H}} | \psi \rangle.
\end{equation}
In Eqs. \ref{d_ham_q} -- \ref{d_schr}, $q_i$ and $p_i$ denote respectively generalized nuclear coordinates and momenta,
$\mathrm{\hat{H}}$ denotes the electronic hamiltonian, which depends parametrically on the nuclear coordinates,
and $| \psi \rangle$ is a time dependent wave function representing the current electronic state.
Small latin indices are used to label nuclear degrees of freedom and range between 1 through $3n$.

Within the DIM+SO approach, the electronic wave function, $| \psi \rangle$, can be expanded using basis set wave
functions of Sec. \ref{interact}, $| \Phi_\alpha \rangle$, also parametrically dependent on nuclear coordinates
and, consequently, on time as well,
\begin{equation}\label{d_psi_exp}
| \psi(t) \rangle = \sum_{\alpha} a_\alpha(t) | \Phi_\alpha(q_i(t)) \rangle,
\end{equation}
with $\alpha$ introduced above, $\alpha = [k,p_m,s_z]$.
The electronic hamiltonian can also be expressed in an expanded form
\begin{equation}\label{d_ham_mtrx}
\mathrm{\hat{H}} = \sum_{\beta,\gamma}\tilde{H}_{\beta\gamma} |\Phi_\beta\rangle\langle\Phi_\gamma|,
\end{equation}
where $\tilde{H}_{\beta\gamma} = S_{\beta\kappa} H_{\kappa\lambda} S_{\lambda\gamma}$ (with
$H_{\kappa\lambda}$
being the DIM+SO hamiltonian matrix given by Eq. \ref{dimso_mtrx}, for simplicity we omit the (SO) upper index), and 
$S_{\alpha\beta} \equiv \langle\Phi_\alpha|\Phi_\beta\rangle$ are overlap matrix elements. Note that matrix
$\tilde{H}_{\beta\gamma}$ is equal to the DIM+SO hamiltonian matrix, $H_{\beta\gamma}$,
if the overlaps are neglected ($S_{\alpha\beta} = 0$ for $\alpha \neq
\beta$) and wavefunctions $|\Phi_\alpha\rangle$ are normalized ($S_{\alpha\alpha} = 1$).\footnote{This is
a usual and sufficiently accurate approximation adopted in all DIM models as yet developed for the rare-gas ionic clusters.} 

After inserting the expanded forms of the electronic hamiltonian and time-dependent electronic wave function
into Eq. \ref{d_ham_p}, one obtains
\begin{equation}\label{d_ham_p_exp}
\dot{p}_i = - \sum_{\alpha,\beta,\gamma,\delta} \left [S_{\alpha\beta}
S_{\gamma\delta} \frac{\partial \tilde{H}_{\beta\gamma}}{\partial q_i} +
D_{\alpha\beta}^{(i)} S_{\gamma\delta} \tilde{H}_{\beta\gamma}
+ S_{\alpha\beta} D_{\delta\gamma}^{(i)*} \tilde{H}_{\beta\gamma}
\right] a_\alpha^* a_\delta,
\end{equation}
where $D_{\alpha\beta}^{(i)} \equiv \langle\Phi_\alpha|\frac{\partial \Phi_\beta}{\partial q_i}\rangle$ are
non-diabatic coupling coefficients, and asterisks denote complex conjugation.
The overlaps and non-diabatic couplings are usually neglected in DIM approaches and,
consequently,  Eq. \ref{d_ham_p_exp}
can be further simplified by
setting $S_{\alpha\beta} \approx \delta_{\alpha\beta}$ and $D_{\alpha\beta}^{(i)} \approx 0$,
(with $\delta_{\alpha\beta}$ being the Kronecker delta),
\begin{equation}\label{d_ham_exp_fin}
\dot{p}_i = - \sum_{\alpha,\beta} a_\alpha^* a_\beta \frac{\partial H_{\alpha\beta}}{\partial q_i},
\end{equation}
where, after neglecting the overlaps, $\tilde{H}_{\alpha\beta}$ is replaced with $H_{\alpha\beta}$.

Further simplification of Eq. \ref{d_ham_exp_fin} is possible if coefficients $a_\alpha$ and matrix elements $H_{\alpha\beta}$,
which are in general complex, are rewritten using their real and imaginary parts,
$a_\alpha = a_\alpha^{(\mathrm{re})} + i a_\alpha^{(\mathrm{im})}$ and
$H_{\alpha\beta} = H_{\alpha\beta}^{(\mathrm{re})} + i H_{\alpha\beta}^{(\mathrm{im})}$,\footnote{ It
directly follows from hermicity of the electronic hamiltonian matrix that its real part,
$H_{\alpha\beta}^{(\mathrm{re})}$, is symmetric and the imaginary part,
$H_{\alpha\beta}^{(\mathrm{im})}$, is antisymmetric. After using this property, we obtain immediately Eq. \ref{d_ham_exp_reim}.}
\begin{equation}\label{d_ham_exp_reim}
\dot{p}_i = - \sum_{\alpha,\beta}\left[ \left (a_\alpha^{(\mathrm{re})} a_\beta^{(\mathrm{re})}
+ a_\alpha^{(\mathrm{im})} a_\beta^{(\mathrm{im})} \right )
\frac{\partial H_{\alpha\beta}^{(\mathrm{re})}}{\partial q_i} +
\left (a_\alpha^{(\mathrm{re})} a_\beta^{(\mathrm{im})} - a_\alpha^{(\mathrm{im})} a_\beta^{(\mathrm{re})} \right )
\frac{\partial H_{\alpha\beta}^{(\mathrm{im})}}{\partial q_i} \right].
\end{equation}
The imaginary part of the electronic hamiltonian matrix is non-zero only if the SO coupling is included.
If it is done using the Cohen-Schneider, {\it atoms-in-molecules} scheme \cite{cohen},
all the imaginary terms are constant as they do not depend on the nuclear positions, and, consequently,
$\frac{\partial H_{\alpha\beta}^{(\mathrm{im})}}{\partial q_i} = 0$. 
The second term on the right-hand-side of Eq. \ref{d_ham_exp_reim}
thus vanishes and the equation can be written in the final
form
\begin{equation}\label{d_ham_exp_re}
\dot{p}_i = - \sum_{\alpha,\beta}\left (a_\alpha^{(\mathrm{re})} a_\beta^{(\mathrm{re})}
+ a_\alpha^{(\mathrm{im})} a_\beta^{(\mathrm{im})} \right )
\frac{\partial H_{\alpha\beta}^{(\mathrm{re})}}{\partial q_i}.
\end{equation}

Similarly, the electronic Schr\"{o}dinger equation, Eq. \ref{d_schr}, can be rewritten
after inserting the expansion of Eq. \ref{d_psi_exp} to
\begin{equation}\label{d_schr_exp}
i\hbar \sum_\beta \dot{a}_\beta |\Phi_\beta\rangle + i\hbar \sum_{\beta,j} a_\beta
\dot{q}_j \frac{\partial|\Phi_\beta\rangle}{\partial q_j} =
\sum_\beta a_\beta \mathrm{\hat{H}}|\Phi_\beta\rangle,
\end{equation}
and after multiplying by $\langle \Phi_\alpha|$ from the left to
\begin{equation}\label{d_schr_exp}
i\hbar \sum_\beta S_{\alpha\beta} \dot{a}_\beta + i\hbar \sum_{\beta,j} D_{\alpha\beta}^{(j)}a_\beta \dot{q}_j =
\sum_{\beta,\gamma,\delta} H_{\gamma\delta} a_\beta.
\end{equation}
A significant simplification of Eq. \ref{d_schr_exp} is further possible if the overlap matrix is replaced with
the Kronecker delta, $S_{\alpha\beta} = \delta_{\alpha\beta}$,
and basis set $| \Phi_\alpha\rangle$ is considered diabatic, $D_{\alpha\beta}^{(i)} = 0$,
\begin{equation}\label{d_schr_exp_noovrlp}
i\hbar \dot{a}_\alpha = \sum_\beta H_{\alpha\beta} a_\beta.
\end{equation}

Eq. \ref{d_schr_exp_noovrlp} must be treated with care, however, namely due to rapid oscillations
occurring in the electronic wave function and, consequently, also in
expansion coefficients $a_\alpha$.
A special scheme has been developed for tackling this problem in the previous work.
Since it is of technical rather than methodological importance, it is not discussed here and the reader
is directed to Ref. \cite{janecek06} for details.
\subsection{Inclusion of quantum decoherence}
As shown elsewhere \cite{janecek09}, quantum decoherence is important, particularly for the heavy rare gases, krypton and xenon.
It is introduced into the mean-field approach by periodically quenching
the electronic wave function.\footnote{Proper settings of the quenching period was thoroughly discussed in \cite{janecek09}. In this work we use
quenching period $t_\mathrm{quench} = 100$\,fs.}
We denote this extended dynamical approach by MFQ (Mean Field with Quenchings).
The quenching algorithm comprises basically two steps.
Firstly, the probabilities for collapsing
the current electronic wavefunction into one of adiabatic states is calculated
and a wave function collapse is proposed according to these probabilities.
Secondly, in case the proposed collapse has been accepted, the kinetic energy of nuclei is adjusted
so that the total energy of the system remains unchanged. The proposed jump can be, in general, rejected in both steps
of the present algorithm
and, in that case, the system resumes the coherent evolution until the next hop attempt.\footnote{For the algorithm used in this work,
AMP (see below), the proposed electronic jump is always accepted in the first step and rejection can occur only during the second step,
namely, if there is not enough kinetic energy to cover expenses of an upward electronic jump.}

Several quenchings schemes have been developed previously \cite{janecek09}. In this work we use
computationally cheap, but several times successfully tested MFQ-AMP/S algorithm.
The procedure starts with calculating the adiabatic amplitudes of the current electronic wave function (hence the acronym AMP).
More specifically, 
the normalized probability for collapsing the current electronic state, $\psi$, to a particular adiabatic state, $\phi_\mu$,
is calculated as
\begin{equation}
\label{AMPprob}
g_{\psi \rightarrow \mu}^\mathrm{AMP}= \rho_{\mu\mu},
\end{equation}
where $\rho_{\mu\mu}$ represents the diagonal element of the electronic density matrix ($\rho_{\mu\nu} \equiv c_\mu c_\nu^*$ and $c_\mu$
are amplitudes of the current electronic wave function, $\psi$, expanded in the adiabatic basis set,
$\psi = \sum_\mu c_\mu \phi_\mu$). After the electronic jump is complete, the kinetic energy of nuclei is adjusted 
so that the total energy of the system is conserved. In the MFQ-AMP/S model, this is achieved by scaling (hence the third acronym, S)
nuclear velocities, as rationalized in \cite{janecek09}.
%
%
%
\section{Radiative dynamics}
\label{rad_dyn}
After the non-radiative dynamics is stopped at time $t_\mathrm{DD}$, each trajectory is evaluated as an ensemble undergoing
first-order decay due to radiative transitions in the electronic subsystem. In principle, many decay processes may occur in such an ensemble,
both parallel and serial, which may lead to a complex system of coupled first-order equations governing the time evolution
of this ensemble. In principle, such equations can be derived and solved.
Nevertheless, since in our case a) transitions are expected only from the upper family of states of the charged fragment,
an excited state resulting for the particular
trajectory from the non-radiative
dynamics at $t_\mathrm{DD}$, to the lower family of states and b) the fragments undergo, after the radiative transition,
a rapid non-radiative decay, the radiative processes can be assumed parallel and a simplified set of decay
equations can be used. In particular, if a population of $n_{I0}$ identical initial states from the upper family of states (e.g., all being state $I$)
is considered for a particular trajectory and assumed to decay to the lower family of states ($J$),
the corresponding population numbers will change with time $\Delta t =t-t_\mathrm{DD}$ according to (dot denotes the time derivative)
\begin{equation}
\label{evol_equ}
\dot{n}_I = -\sum_J \Gamma_{IJ} n_I, \quad \dot{n}_J = \Gamma_{IJ} n_I,
\end{equation}
with initial conditions $n_I(\Delta t=0) = n_{I0}$ (=1 for one particular trajectory) and $n_J(\Delta t=0) = 0$. It is easy to find, that the only solution to these equations is given
by Eq. (2) of the letter, namely,
\begin{equation}
\label{occup_nmbs}
n_J(\Delta t) = n_{I0} (1-e^{-\Gamma \Delta t}) \Gamma_{IJ}/\Gamma, \quad 
n_I(\Delta t) = n_{I0} e^{-\Gamma \Delta t},
\end{equation}
where $\Gamma = \sum_{K = 1}^{I-1} \Gamma_{IK}$ and $\Delta t \approx t$ since $t \gg t_\mathrm{DD}$.

Note also that $n_J(\Delta t)\vert_{n_{I0}=1}$ gives the probability that, at time $\Delta t$, the system will be found
in state $J$, and $n_I( \Delta t)\vert_{n_{I0}=1}$ is the probability of surviving the system in excited state $I$.
The evaluation of fragments at time $\Delta t$ consists then in a cycle repeated for all trajectories and comprising the
following steps:
\begin{enumerate}
\item{identify the fragmentation channel corresponding to the particular trajectory,}
\item{subtract from the total number of trajectories leading to the same fragmentation channel at $t_\mathrm{DD}$ value of $1 - n_I(\Delta t)\vert_{n_{I0}=1}$,}
\item{identify the fragmentation channel for each state $J$ (this can be done either by running additional non-radiative dynamical simulation
or by simple energetic considerations),}
\item{add $n_J(\Delta t)\vert_{n_{I0}=1}$ to the number of trajectories leading at $t_\mathrm{DD}$ to the same fragmentation channel.}
\end{enumerate}
After this cycle is complete, one gets updated abundances of fragments as should be detected at time $\Delta t \approx t$ of radiative decay.

The decay rates of Eq. \ref{evol_equ} are calculated from a standard formula for spontaneous radiation (Eq. 1 of the letter),
\begin{equation}
\label{gamma}
\Gamma_{IJ} = \frac{1}{{3\pi \varepsilon _0 \hbar ^4 c^3 }}\left( {E_I  - E_J } \right)^3 \left| {\mu _{IJ} } \right|^2,
\end{equation}
where the transition dipole moment is obtained for a particular charged fragment geometry, ${\bf R}$,
within the point-charge approximation \cite{ikegami93},
\begin{equation} \label{tdm}
{\mu}_{IJ}({\bf R})\approx e \sum_{k=1}^n\sum_{m=x}^z\sum_{s_z = -1/2}^{+1/2}{c_{kp_ms_z}^{(I)}}^{ \ast}c_{kp_ms_z}^{(J)}{\bf R}_k.
\end{equation}
The first sum of Eq, \ref{tdm} runs over all atoms in the charged fragment and $c_{k p_m s_z}^{(I)}$ and $c_{k p_m s_z}^{(J)}$ are amplitudes
of adiabatic states $I$ and $J$, respectively, expressed in the DIM+SO basis set introduced in Sec. \ref{interact}.
Alike in our earlier work, the point-charge approximation has been further improved in the present work by including
damped polarization effects \cite{naumkin00} consisting in a replacement
\begin{equation} \label{tdmid}
{\bf R}_k \rightarrow {\bf R}_k \sum_{i \neq k} \alpha^*_{\mathrm{eff}}(R_{ik})\frac{{\bf R}_{ik}}{{R_{ik}}^3},
\end{equation}
where ${\bf R}_{ik} = {\bf R}_{i}-{\bf R}_{j}$, $R_{ik}=|{\bf R}_{ik}|$, and $\alpha^*_{\mathrm{eff}}(R_{ik})$ is a damped
effective polarizibility expressed in atomic units \cite{naumkin00},
\begin{equation} \label{alfa}
\alpha_{\mathrm{eff}}^*(R) = \frac{N_e}{(\sqrt{N_e/\alpha^*}+1/R )^2}.
\end{equation}

%
%
%
\bibliography{refs}
\bibliographystyle{unsrt}

\end{document}